\documentclass[]{interact}

\usepackage{epstopdf}% To incorporate .eps illustrations using PDFLaTeX, etc.
\usepackage[caption=false]{subfig}% Support for small, `sub' figures and tables
\usepackage[natbibapa,nodoi]{apacite}
\theoremstyle{plain}% Theorem-like structures provided by amsthm.sty

\theoremstyle{definition}

\theoremstyle{remark}

\begin{document}

% \articletype{ORIGINAL RESEARCH ARTICLES}% Specify the article type or omit as appropriate

\title{\textbf{Communication Platform for Non-verbal Autistic children in Oman using Android mobile}}

% \maketitle
\author{
\name{Amna Al-Araimi and Yue Zheng\textsuperscript{*}\thanks{Corresponding author: Email: yz15u22@soton.ac.uk } and Haiming Liu\textsuperscript{*}\thanks{Corresponding author: Email: h.liu@soton.ac.uk}}
\affil{School of Electronics and Computer Science, University of Southampton, Hampshire, United Kingdom}
}

\makeatletter
\newcommand{\@presented}[1]{\textit{#1}}
\makeatother
\maketitle

\begin{abstract}
This paper discuss about the issue regarding Non-verbal Autism Spectrum Disorder. It has been seen that this mental disorder is listed in major parts of the world including US, UK and India. To mitigate this type of disorder a wide range of smartphone, computers and artificial intelligence has been used. This technology has helped the population to cope up with socialization and the communication needs. Many application regarding this technology has been developed specializing communication capabilities of the non–verbal autistic children. This thesis project has given a proposal for developing a platform that included a web panel and an android mobile application together which helped the non-verbal autistic children in communication process and it has helped them to express themselves specially in Oman. Different intervention in this case have been merged for improving life of the people’s spectrum from autism. The main problem has been assessed in this case is that the implemented fragmented approach is not suitable for the autistic child. The augmented reality framework has the capability of poking the kid’s autism toward increment in creative play. This framework will give the child an opportunity to view themselves on the PC screen like the mirrors.
\end{abstract}


\section{Introduction}
\subsection{Background}
Autism Spectrum Disorder (ASD) is a neurodevelopmental condition that affects children's communication and behavior, typically emerging within the first two years of life. Globally, around 1\% of the population is affected by ASD \citep{talantseva2023global}, presenting significant challenges in communication and social interaction, which can make independent living difficult. Children with ASD often struggle to form social relationships, which can be isolating for both the children and their caregivers \citep{kuriakose2016design}.

Although various tools, such as voice communication aids and Content-Based Image Retrieval (CBIR) systems, are available to assist non-verbal individuals with ASD, these systems often lack integration\citep{teevan2005personalizing}. Voice communication aids are crucial but suffer from compatibility issues across different platforms\citep{shminan2017autipecs}, while CBIR systems, though intuitive, do not fully address the diverse needs of individuals with ASD. Additionally, the absence of multilingual support, especially for languages like Arabic, further limits the accessibility and effectiveness of these tools for children in regions like Oman.

To address these challenges, we aim to design a communication platform that helps non-verbal autistic children in Oman better engage with the world and express themselves. This platform will integrate existing tools and provide multilingual support, offering a more cohesive solution to enhance communication for these children while also supporting their families and therapists.

\subsection{Problem Statement and Literature Review}
Non-verbal children with ASD face significant communication challenges that hinder their ability to express needs and emotions, which in turn exacerbates their isolation and makes social integration difficult \citep{reyes2020emotion}. This lack of communication also prevents them from recognizing available support, increasing their vulnerability to misunderstandings. Additionally, ASD children often rely heavily on their parents for daily care, but many parents struggle to accurately interpret their child’s emotional and intellectual cues, complicating progress tracking and early intervention \citep{depape2015parents}.There is also a critical need for expert assessments to diagnose specific issues and guide evidence-based interventions \citep{odom2010evidence}. However, the fragmentation of available tools, lack of platform integration, and insufficient language support, particularly in regions like Oman, further complicate the effectiveness of these efforts.

Enhancements in CBIR technology can significantly support non-verbal autistic children. \cite{liu2009four} proposed integrating CBIR with voice communication and text-based systems to enhance multi-modal interaction, offering a more comprehensive communication platform. \cite{kodgirwar2016application} emphasized that CBIR allows systems to retrieve data based on graphical searches, which is particularly effective for autistic children, who process visuals better than words \citep{kamaruzaman2014form}. Improvements, such as enabling sketch-based queries, make CBIR even more intuitive for children with autism to interact with educational content \citep{agarwal2016sketch}. To be effective, such an integrated platform must be visually rich and feature multi-modal interactions, combining visual and auditory elements to support personalized learning and communication. The subsequent sections will outline the methodology used to develop and assess such a system.

\section{METHODOLOGY}
\subsection{Data Collection}

To better understand the specific challenges faced by non-verbal autistic children and their caregivers, we employed a combination of surveys, interviews, and case studies. Surveys were conducted with parents, therapists, and educators to gather information on the effectiveness of current communication tools, the challenges of implementing them, and the particular needs of Arabic-speaking children in Oman. Additionally, semi-structured interviews with therapists and specialists working directly with non-verbal autistic children provided deeper insights into the daily challenges faced by children, parents, and professionals when using existing tools. Several case studies were also conducted in institutions in Oman, observing the interactions between non-verbal autistic children and their therapists. These observations highlighted key issues such as communication delays, difficulty accessing appropriate tools, and the heavy reliance on caregivers for daily communication support. Table \ref{paticipant_info} provides a summary of the participants involved in the study.
\begin{table}[h!]
\centering
\caption{Participant Information}
\label{paticipant_info}
\renewcommand{\arraystretch}{1.2}
\begin{tabular}{p{2cm}p{2.5cm}p{4cm}p{2.5cm}p{1cm}}
\toprule
\textbf{Institution} & \textbf{Therapist} & \textbf{Participant number} & \textbf{Gender} & \textbf{Age} \\ 
\midrule
1 & A & 1 & Male & 7 \\ 
  & A & 2 & Male & 7 \\ 
  & B & 3 & Male & 10 \\ 
  & C & 4 & Female & 5 \\ 
\midrule
2 & D & 5 & Male & 8 \\ 
  & D  & 6 & Male & 11 \\ 
  & E & 7 & Male & 9 \\ 
\midrule
3 & F & 8 & Female & 11 \\ 
  & G & 9 & Female & 6 \\ 
\midrule
4 & AA & 10 & Male & 4 \\ 
  & BB & 11 & Female & 10 \\ 
\bottomrule
\end{tabular}
\end{table}
\subsection{Design Development}
Following data collection, the development of the communication platform followed an Iterative Prototyping Methodology \citep{camburn2015systematic}, allowing for continuous refinement based on user feedback. Initial requirements were gathered through surveys, interviews, and case studies, identifying key needs such as multi-language support (specifically Arabic), voice communication features, and CBIR-based visual search functionality. A prototype was then developed, integrating voice communication aids and CBIR technology, and tested by therapists and parents to evaluate its usability and effectiveness for non-verbal autistic children. Subsequent real-world testing provided feedback from children and caregivers, leading to several rounds of refinement to enhance the platform’s functionality and user experience. Once the platform met performance criteria, it was implemented in selected institutions in Oman, with training provided to therapists and parents. Regular follow-up evaluations ensured the platform’s continued effectiveness in supporting the children’s communication needs.

\begin{figure}[h!]
\centering
\includegraphics[width=1\textwidth]{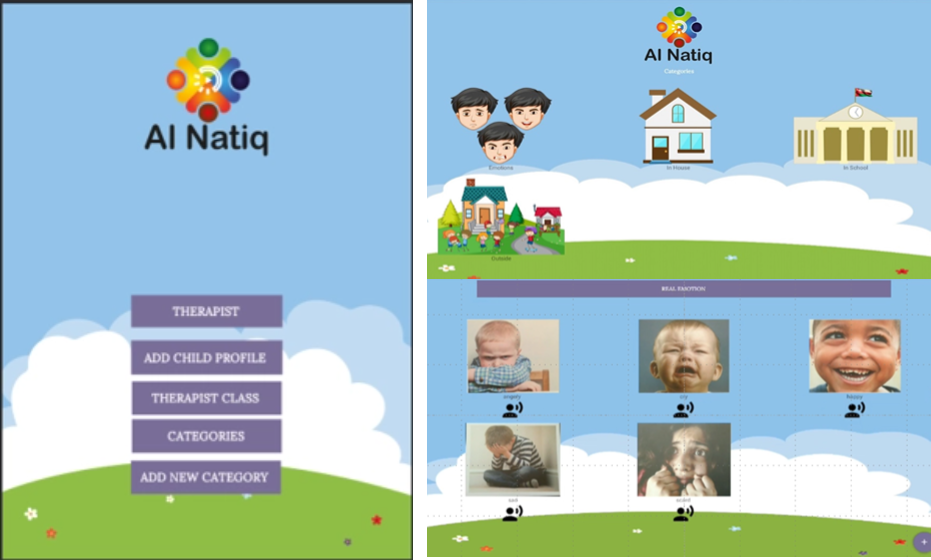} % 调整图片宽度
\caption{Application Interface Demonstration}
\label{fig:app_interface_demo} % 添加标签，用于引用
\end{figure}

Our application, referencing Figure \ref{fig:app_interface_demo}, is designed primarily for therapists, allowing them to create and manage individual child profiles. Therapists can customize sessions by adding various categories to tailor their approach to each child's needs. While the application includes four pre-built categories, users have the flexibility to upload diverse types of images with corresponding descriptions. This feature is particularly valuable as it accommodates regional variations in educational materials, ensuring that the platform can be adapted to different cultural and linguistic contexts.

\section{Findings of the experiments}
This section presents the outcomes derived from the initial design, development, and evaluation of the Android-based communication platform tailored for non-verbal autistic children in Oman. In addition to limited observational trials with end users, the study included qualitative input from stakeholders, such as speech-language therapists, caregivers, and special education specialists. Examining the system's usability, cultural suitability, and communication effectiveness from a variety of angles is the aim of the findings.

\subsection{Therapist and Expert Feedback}

Three to five speech-language therapists (SLTs) and special educators who work with non-verbal children ages four to eleven participated in the participatory design process, which included organized conversations and practical demonstrations. The overall favorable response to the application's image-voice pairing interface was a noteworthy result. Therapists observed that children's involvement and cognitive obstacles were much reduced when culturally appropriate visual stimuli were used. Crucially, the Arabic voice feedback, which is sometimes absent from popular AAC programs, was regarded as a significant breakthrough.

Participants emphasized that the structured yet flexible design of the application made it easier to personalize content based on each child's developmental stage and expressive capacity. Furthermore, the simplicity of the UI, combined with the option for caregivers to add custom images or record familiar voices, promoted a sense of familiarity and emotional safety key considerations when working with children with ASD.

\subsection{Usability with Children}

Initial trials were conducted with a limited number of non-verbal autistic children under the supervision of their therapists and caregivers. While a full-scale empirical usability study remains a task for future work, early observations suggest that the app effectively facilitates basic emotional and needs-based expression. For example, when prompted with images associated with feelings (e.g., happy, angry, scared), children were able to use the app’s touch interface to select appropriate representations. In several cases, children who had previously struggled with face-to-face interaction were observed gradually becoming more responsive when engaging with the app, particularly when voice feedback was paired with visually distinct icons.

A noteworthy observation included a 10-year-old girl with severe ASD and a facial scar that inhibited her willingness to engage in social interaction. Initially withdrawn and avoiding eye contact, she eventually began participating in tasks by quickly tapping the correct emotional expression images after minimal prompting. With repeated interaction, the child became increasingly confident and communicative during the session, even managing a brief handshake at the end. This anecdotal case underscores the potential of image-voice platforms to scaffold social participation and communicative growth in real-time.

\subsection{Content-Based Image Retrieval Module Feasibility}

The integration of a CBIR framework was assessed as a forward-looking feature designed to support personalized content updates. Although not fully deployed in this prototype version, expert interviews highlighted the need for dynamic content suggestions that adapt to the user’s interaction history and evolving needs. The feasibility of CBIR within the Omani context is particularly promising, considering the visual richness of local culture and the limited availability of Arabic-language content in existing systems.

Therapists expressed interest in a potential CBIR enhanced version that could suggest contextually appropriate icons or images based on prior usage or caregiver input. For instance, if a child frequently uses images related to school or outdoor activities, the system could learn to prioritize similar or complementary items. This adaptive mechanism was seen as essential in transitioning from static communication tools to more intelligent, responsive systems.

\subsection{Cultural and Linguistic Adaptation}

A recurrent theme across all feedback was the inadequacy of existing AAC tools in supporting Arabic language nuances and cultural content. Most available systems are either Western-centric in their design or limited in their multilingual adaptability. The prototype presented in this study was praised for its deliberate effort to incorporate Arabic voice feedback, culturally familiar icons, and region-specific references.

Participants emphasized that effective communication tools for children with ASD must go beyond linguistic translation. They must also embed cultural practices, everyday scenarios, and social expectations that align with the child's lived environment. In this respect, the current application filled an important gap and served as a culturally grounded alternative to imported solutions.

\section{Conclusion and Future Work}
This study introduced the design and preliminary evaluation of a mobile communication platform tailored for non-verbal autistic children in Oman. The application was developed to address a clear gap in existing assistive technologies, particularly the lack of culturally and linguistically appropriate tools for Arabic-speaking children with ASD. Through collaboration with therapists and caregivers, the system incorporated regionally relevant imagery and Arabic voice feedback to create a more accessible and engaging environment for communication.

The findings demonstrated that combining voice feedback with intuitive visual interfaces can significantly improve engagement and communication ability among non-verbal children. Therapists observed that children, even those initially reluctant to participate, gradually showed more responsiveness and confidence when interacting with the app. These observations highlight the importance of designing tools that are not only functionally supportive but also emotionally safe and culturally aligned.

The platform’s current version successfully supports basic communication tasks such as expressing emotions or identifying needs through image selection. However, the system remains limited in its adaptability and personalization. While users can customize content manually, the lack of intelligent content recommendation restricts the platform’s scalability and responsiveness to user behavior.

Looking forward, several directions will be pursued to enhance the system's effectiveness and relevance. The first is the integration of CBIR algorithms, allowing the system to dynamically suggest relevant images based on user interactions. This will help the application evolve with the user, offering a more personalized experience.

The second direction involves extending the platform’s modality support. Future versions will incorporate gesture recognition and sign language features, thereby accommodating a broader range of communicative expressions. These enhancements will be grounded in ongoing research into multimodal inclusive information access, which is the central focus of the author’s doctoral work.

Additionally, further experiments with a larger and more diverse group of participants are planned. These will include both behavioral observations and physiological data collection to evaluate the cognitive load and emotional responses elicited by the system. The insights gained will guide iterative improvements and contribute to the broader understanding of how inclusive technologies can support individuals with disabilities.

In conclusion, this work represents an early but important step towards creating inclusive, culturally sensitive, and multimodal communication tools for non-verbal autistic children. By aligning the system design with user needs, regional context, and long-term academic goals, the research lays a foundation for a more inclusive future in assistive technology and information retrieval.

\bibliographystyle{apacite}
\bibliography{main}

\end{document}